\documentclass[12pt]{article}
\def\a{\alpha}

\def\g{\gamma}
\def\d{\delta}

\def\e{\epsilon}

\def\m{\mu}

\def\p{\psi}
\def\o{\omega}

\begin{document}
\renewcommand{\thefootnote}{\fnsymbol{footnote}}

\setcounter{page}{1}

\begin{center}
{\Large{\bf Reply to Professor Peter van Nieuwenhuizen in connection with 
hep-th/0408137\\}}
\vspace{1cm}
{\bf Vyacheslav A. Soroka\footnote{E-mail: vsoroka@kipt.kharkov.ua}}
\vspace{1cm}\\
{\it Institute for Theoretical Physics, NSC ``Kharkov Institute of Physics and 
Technology'' Akademicheskaya Street 1, 61108 Kharkov, Ukraine}\\
\vspace{1.5cm}
\end{center}

Professor Peter van Nieuwenhuizen in hep-th/0408137 wrote:
\begin{quote}
``In an early article, Volkov and Soroka \cite{vs} gauged the super Poincar\'e 
algebra (not the super anti de Sitter algebra), but did not prove its 
invariance under supersymmetry. They used a first-order formalism and found 
$\d\o=0$. This agrees with (21)
$$
{\d\o_\m}^{mn}=\a\bar\e\g^{mn}\p_\m
\eqno{(21)}
$$
in the limit $\a=0$. As we have explained, this result is incorrect. They 
implicitly assumed that their action would be invariant, and concluded that
supergravities exist for any $N$. Careful study, using $\d\o\ne0$, shows that
$N\le8$.''
\end{quote}

In \cite{vs} we considered the spontaneously broken $N$-extended super 
Poincar\'e gauge group and with the use of the Cartan invariant forms 
constructed five invariants containing respectively the action for 
the graviton, the kinetic term for the gravitinos, the mass term of the 
gravitinos, the cosmological term and the term for the Yang-Mills fields.

The term ``incorrect'' concerning something, as I understand, means that this 
something contains the mistakes. But our papers under the above mentioned 
statement of a question do not contain any mistake and therefore are 
completely correct.

The supersymmetry transformation low $\d\o=0$ in our papers is a consequence
of the group structure and is also completely correct.

In accordance with the Yang-Mills prescription, for the $N$-extended super
Poincar\'e gauge group we use the adjoint representation, which has no 
restriction $N\le8$.

The statement ``...but did not prove its invariance under supersymmetry''
indicates that the author of hep-th/0408137 did not familiarize yourself with 
the Cartan method of the invariant forms.

By concluding, I believe that the assertion, that our papers are incorrect,
misleads the scientific community, especially young physicists.


\begin{thebibliography}{999}
\bibitem{vs}D. Volkov and V. Soroka, {\it JETP Lett.} {\bf18} (1973) 312 and
{\it Theor. Math. Phys.} {\bf20} (1974) 829.

\end{thebibliography}
\end{document}